\documentclass[aip,reprint]{revtex4-1}
\usepackage{graphicx}
\usepackage{epsfig}
\usepackage{color}
\usepackage{epstopdf}
\usepackage{amsmath}
\usepackage{caption}
\usepackage{subcaption}

\newcommand{\grl}{    {Geophys. Res. Lett.}}
\newcommand{\jgr}{    {J. Geophys. Res.}}
\newcommand{\ssr}{    {Space Sci. Rev.}}

\newcommand{\apjo}{ {Astrophys. J. }}

\newcommand{\pree}{ {Phys. Rev. E }}
\newcommand{\mnras}{ {Monthly Notices of the Royal Astronomical Society}}
\newcommand{\natu}{    {Nature}}

\begin{document}


\title{Mapping for nonlinear electron interaction with whistler-mode waves.} 



\author{A. V. Artemyev}
 \altaffiliation[Also at] { Space Research Institute, RAS, Moscow, Russia}
 \email{aartemyev@igpp.ucla.edu}
\affiliation{
Institute of Geophysics and Planetary Physics, University of California, Los Angeles, USA.
}%

\author{A. I. Neishtadt}
 \altaffiliation[Also at]  { Space Research Institute, RAS, Moscow, Russia}
\affiliation{
Department of Mathematical Sciences, Loughborough University, Loughborough LE11 3TU,
United Kingdom.
}%

\author{A. A. Vasiliev}
\affiliation{
Space Research Institute, RAS, Moscow, Russia.
}%

\date{\today}

\begin{abstract}
The resonant interaction of relativistic electrons and whistler waves is an important mechanism of electron acceleration and scattering in the Earth radiation belts and other space plasma systems. For low amplitude waves, such an interaction is well described by the quasi-linear diffusion theory, whereas nonlinear resonant effects induced by high-amplitude waves are mostly investigated (analytically and numerically) using the test particle approach. In this paper, we develop a mapping technique for the description of this nonlinear resonant interaction. Using the Hamiltonian theory for resonant systems, we derive the main characteristics of electron transport in the phase space and combine these characteristics to construct the map. This map can be considered as a generalization of the classical Chirikov map for systems with nondiffusive particle transport and allows us to model the long-term evolution of the electron distribution function.
\end{abstract}

\pacs{}

\maketitle

\section{Introduction}
Whistler waves are electromagnetic emissions within the frequency range from the lower-hybrid up to electron cyclotron frequency widely observed in space \cite{Li11, Agapitov13:jgr, Wilson12, Tong19:ApJ, Zhang18:whistlers&injections} and laboratory \cite{Stenzel99, VanCompernolle15} plasmas. These waves are generated by various types of electron distributions with thermal anisotropy \cite{Vedenov&Sagdeev61, Kennel&Wong67},  beam distributions \cite{Krafft00,Volokitin&Krafft01:lhd, Volokitin&Krafft01:whistlers, An16}, or both \cite{Artemyev16:ssr, Li16, Mourenas15}, and they play an important role in the isotropisation of originally unstable electron distributions \cite{Gary&Wang96, Tao17:saturation, Kuzichev19}. A classical theory of whistler wave resonant interaction with electrons is the quasi-linear theory \cite{Vedenov62, Kennel&Engelmann66, Lerche68} that assumes a broad spectrum of low amplitude waves. This theory allows to describe the main characteristics of electron acceleration \cite{Thorne13:nature, Li13} and scattering \cite{Thorne10:Nature, Ni16:ssr} in the Earth radiation belts, in the solar wind \cite{Shaaban19}, and at the Earth bow shock \cite{Veltri92, Veltri&Zimbardo93}. However, quasi-linear theory cannot describe resonant interactions with the very intense coherent waves \cite{Karpman74:ssr, Tao12} often observed in space plasmas \cite{Titova03, Agapitov14:jgr:acceleration, Cattell11:Wilson, Wilson12, Zhang18:jgr:intensewaves, Zhang19:grl, Tyler19}. Such sufficiently intense whistlers can resonate nonlinearly with electrons \cite{Nunn71, Karpman74, Bell84}. Such nonlinear interaction can lead to phase trapping or non-diffusive scattering of particles \cite{Solovev&Shkliar86, Albert93} and can result in a very fast electron acceleration \cite{Albert02, Demekhov06, Demekhov09, Omura07, Furuya08, Tao12:GRL, Yoon13:whistlers}. Therefore, the effects of nonlinear resonant interaction are actively investigated (see reviews in Refs. \onlinecite{Omura91:review, Shklyar09:review, Albert13:AGU, Artemyev16:ssr, Artemyev18:cnsns}).

Since self-consistent Vlasov or Particle-In-Cell simulations of whistler wave generation and their resonances with electrons \cite{Katoh&Omura07, Fu14:radiation_belts, Tao17:generation, Kuzichev19:pop} can hardly cover the long-term dynamics of the electron distribution in realistic space plasma systems, alternative approaches need to be considered. Beside the test particle approach (i.e., the numerical integration of a large number of electron trajectories \cite{Nunn&Omura15, Agapitov15:grl:acceleration, Kitahara&Katoh19}), the most interesting approach for the investigation of nonlinear electron resonances with whistler waves consists in the derivation of a kinetic equation (master-equation \cite{book:VanKampen03}) describing the evolution of the electron distribution. This approach generalizes the quasi-linear diffusion equation by including terms responsible for electron nonlinear acceleration and scattering \cite{Furuya08, Artemyev14:grl:fast_transport, Omura15, Artemyev16:pop:letter}. Such terms can be derived analytically \cite{Artemyev17:pre, Vainchtein18:jgr} or numerically \cite{Omura15, Hsieh&Omura17, Hsieh&Omura17:radio_science}.

A less investigated but potentially useful approach is the mapping technique already widely used for systems with small wave amplitudes \cite{Lichtenberg&Lieberman83:book, bookSagdeev88}. The well-known Chirikov map \cite{Chirikov79} describes phase space diffusion and transport induced by periodical random jumps of particle momentum. The resonance of electrons and whistler waves results in a similar type of dynamics: each resonant interaction corresponds to an electron energy (and pitch-angle) jump inducing particle transport in phase space. For small amplitude whistler waves, the map of electron resonant jumps is quite similar to the Chirikov map \cite{Khazanov13, Khazanov14}, but nonlinear resonant interaction should significantly change such a map. In this study, we  develop a map describing electron motion in a system with multiple passages through nonlinear resonances. We have also demonstrated that this map models well the electron distribution evolution and can be used to study the radiation belt dynamics.

\section{Basic equations}
To derive the basic properties of the nonlinear electron ($m_e$ is the rest mass and $-e$ is the charge) interaction with field-aligned whistler waves ($\omega$ is a constant frequency, $k(\omega,s)$ is the wave number given by the cold plasma dispersion relation \cite{bookStix62} and depending on the field-aligned coordinate $s$) we consider Hamiltonian (see details in, e.g., Refs. \onlinecite{Artemyev18:jpp, Vainchtein18:jgr}):
\begin{eqnarray}
H &=& m_e c^2 \gamma  + U_w \left( {s,I_x } \right)\sin \left( {\phi  + \psi } \right)\nonumber\\ \gamma  &=& \sqrt {1 + \frac{{p_\parallel ^2 }}{{m_e^2 c^2 }} + \frac{{2I_x \Omega _{ce} }}{{m_e c^2 }}}
\label{eq01}
\end{eqnarray}
where $c$ is the speed of light, $\gamma$ is the gamma factor of the gyroaveraged system, $\Omega_{ce}=eB_0(s)/m_ec$ is the electron gyrofrequency in the background magnetic field $B_0(s)$ given by the reduced dipole model \cite{Bell84}, $U_w=\sqrt{2I_x\Omega_{ce}m_e}eB_w/\gamma m_e ck$ with $B_w$ the wave amplitude, $I_x$ is the magnetic moment normalized in a such a way that $I_x\Omega_{ce}$ has the dimension of energy. The conjugate pairs of variables in Eq. (\ref{eq01}) are field-aligned coordinate and momentum, $(s,p_\parallel)$, and gyrophase and magnetic moment, $\psi, I_x$. Wave phase $\phi$ is given by the differential equation: $\dot\phi=k(s)\dot s-\omega$ (we omit the argument $\omega$ in the function $k$). In system (\ref{eq01}) phases $\phi$, $\psi$ change much faster than variables $s$, $p_\parallel$, $I_x$ because wave frequency $\omega$ and gyrofrequency $\Omega_{ce}$ are much larger than the electron bounce frequency $\sim c/R$ where $R$ is a spatial scale of $B_0(s)$ gradient. Therefore, the first small parameter of the system is $c/R\Omega_{ce} \ll 1$. The second small parameter of the system is $B_w/B_0 \ll 1$, and we consider sufficiently intense waves with $B_w/B_0 \geq c/R\Omega_{ce}$ (this condition is satisfied for a significant fraction of whistler waves observed in the Earth radiation belts, see Ref. \onlinecite{Zhang18:jgr:intensewaves}).

\begin{figure*}
\includegraphics[width=0.95\textwidth]{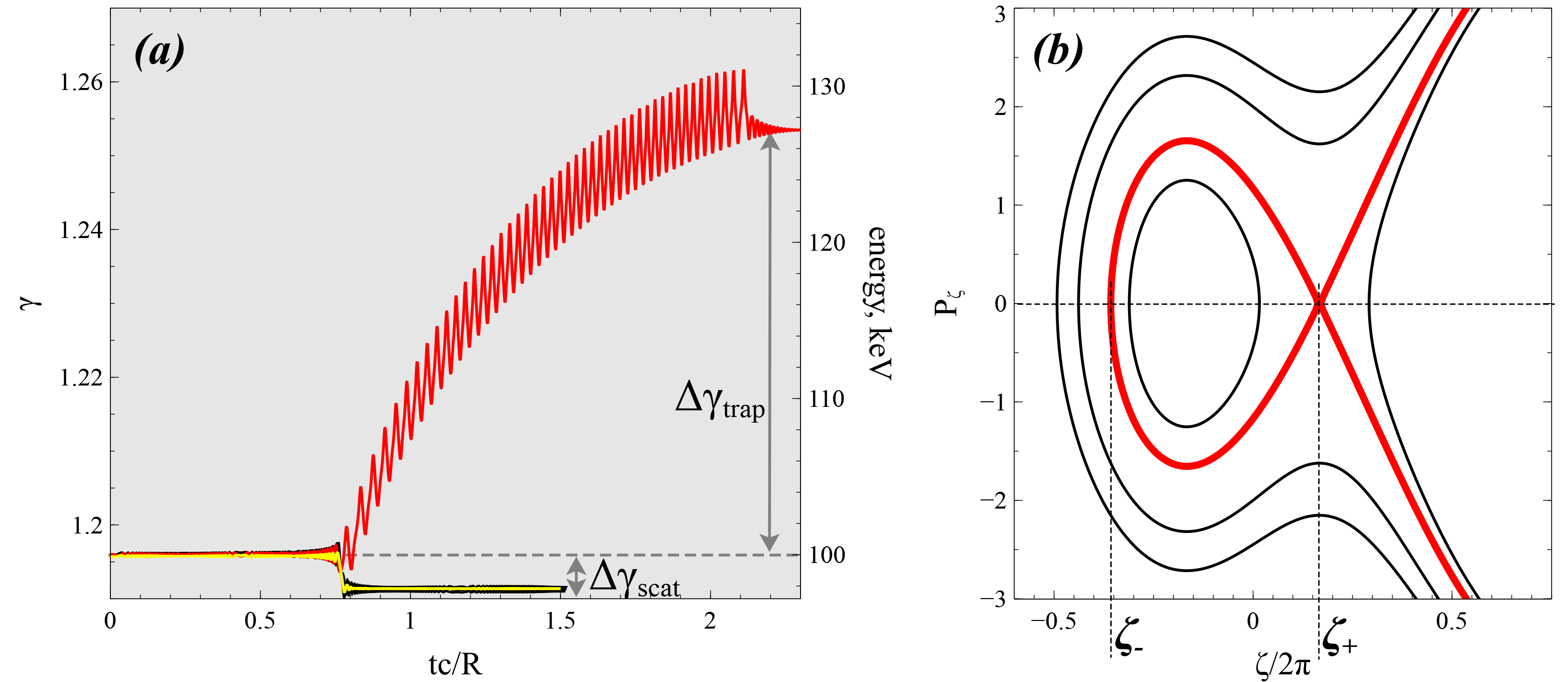}
\caption{(a) Change of electron energy due scattering (black; yellow shows averaged energy of scattered particle) and trapping (red). The time interval of one resonant interaction is shown. All electrons have initially the same energy and pitch-angle. For these trajectories and throughout the paper we consider a curvature-free dipole magnetic field \citep{Bell84} with the radial distance from the Earth $R=4.5$ of the Earth radii. The whistler wave frequency is $0.35$ times the electron cyclotron frequency at the equator, and plasma frequency equals to $4.5$ of the electron cyclotron frequency at the equator. To evaluate the wave number $k$ we use the cold plasma dispersion of whistler waves \cite{bookStix62}. Wave amplitude is $300$ pT, typical for intense whistlers observed in the radiation belts \cite{Cattell08, Cattell11:Wilson, Zhang19:grl, Tyler19}. The distribution of the wave amplitude along magnetic field lines, $B_w(s)$, is modeled by function $\tanh((\lambda/\delta\lambda_1)^2)\exp(-(\lambda/\delta\lambda_2)^2)$ with $\lambda$ the magnetic latitude ($ds=Rd\lambda\sqrt{1+\sin^2\lambda}\cos\lambda$) and $\delta\lambda_1=2^\circ$, $\delta\lambda_2=20^\circ$. This function fits the observed whistler wave intensity distribution \cite{Agapitov13:jgr}. To simplify the simulations, we consider waves in only one hemisphere, $B_w=0$ for $s<0$, and thus there is only one resonance for electrons within one bounce period. Waves are moving away from the equatorial plane, $s=0$, to large $s$, i.e. only $k>0$ are included. (b) Phase portrait of Hamiltonian $\tilde{H}_I-\Lambda=gP_\zeta^2/2-r\zeta+U_w\sin\zeta$ for $|U_w/r|>1$. \label{fig1}}
\end{figure*}

Figure \ref{fig1}(a) shows several fragments of electron trajectories around the resonance $\dot\phi+\dot\psi=0$ (with $\dot\psi=\partial H/\partial I_x\approx\Omega_{ce}/\gamma$): there are two main effects\cite{Neishtadt14:rms} -- electron trapping into resonance with the energy increase $\Delta\gamma_{trap}$, and electron scattering on the resonance with the energy decrease $\Delta\gamma_{scat}$. We plan to construct a map describing the long-term evolution produced by these two processes.

\begin{figure*}
\includegraphics[width=0.95\textwidth]{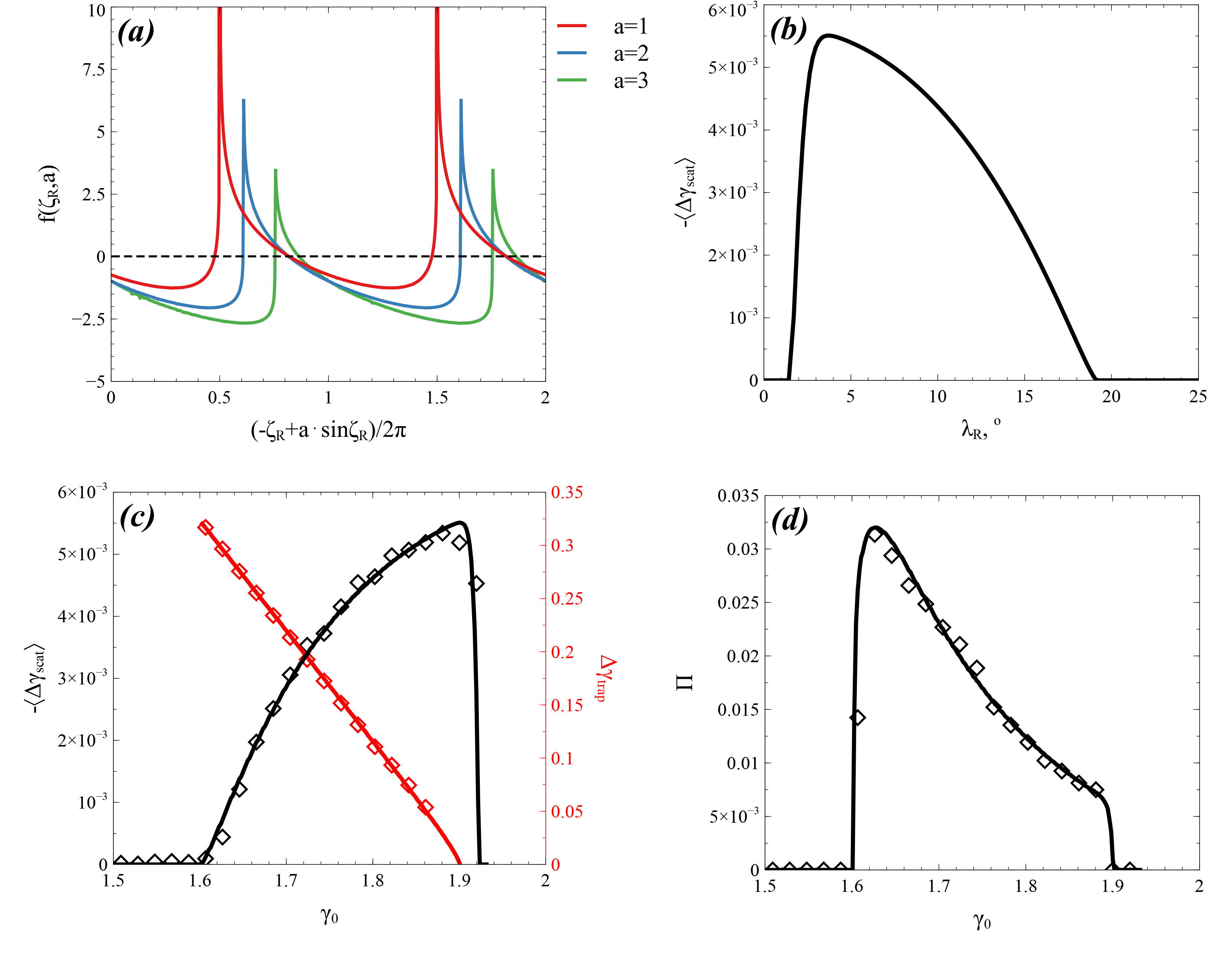}
\caption{Main parameters of the resonant system: (a) function $f(\zeta_R, a)$; (b) function $\langle \Delta\gamma_{scat}(\lambda_R) \rangle$, (c) function $\langle \Delta\gamma_{scat}(\gamma_0) \rangle$; (d) probability of trapping $\Pi(\gamma_0)$. Symbols show the results of numerical simulations ($h/m_ec^2=1.454$ corresponds to, e.g., $\alpha_0=45^\circ$ of the equatorial pitch-angle for $300$ keV electron energy; see details of model parameters in the caption of Fig. \ref{fig1}). \label{fig2} }
\end{figure*}

We start with the determination of $\Delta\gamma_{trap}$, $\Delta\gamma_{scat}$ and their dependencies on particle characteristics. First, we use the generating function $W_1=((\int{k(\tilde{s})d\tilde{s}}-\omega t)+\psi)I+sP$ to introduce phase $\varphi=\phi+\psi$ and conjugate momentum $I$:
\begin{eqnarray}
 H_I  &=&  - \omega I + m_e c^2 \gamma  + U_w \left( {s,I} \right)\sin \varphi  \nonumber\\
 \gamma  &=& \sqrt {1 + \frac{{\left( {P + kI} \right)^2 }}{{m_e^2 c^2 }} + \frac{{2 { I}\Omega _{ce} }}{{m_e c^2 }}}  \label{eq02}
\end{eqnarray}
Pairs of conjugate variables are $(I,\varphi)$ and $(s,P)$. The resonance condition ($\dot\varphi=\partial H_I/\partial I=0$) for Hamiltonian (\ref{eq02}) defines the resonant momentum $I_R(s,P)$:
\begin{eqnarray}
\frac{{kI_R }}{{m_e c}} &=&\sqrt {\frac{{1 - \left( {\Omega _{ce} /kc} \right)^2  - 2\left( {\Omega _{ce} P/km_e c^2 } \right)}}{{\left( {kc/\omega } \right)^2  - 1}}} \nonumber\\
&-& \frac{{\Omega _{ce} }}{{kc}} - \frac{P}{{m_e c}}
\label{eq03}
\end{eqnarray}
We expand Hamiltonian (\ref{eq02}) around the resonance
\begin{eqnarray}
H_I  &\approx& \Lambda  + \frac{1}{2}g\left( {I - I_R } \right)^2  + U_w \left( {s,I_R } \right)\sin \varphi  \label{eq04}
\\
 \Lambda & =&  - \omega I_R  + m_e c^2 \gamma _R ,\quad \gamma _R  = \sqrt {1 + \frac{{\left( {P + kI_R } \right)^2 }}{{m_e^2 c^2 }} + \frac{{2I_R \Omega _{ce} }}{{m_e c^2 }}}  \nonumber\\
 g &=& m_ec^2\left. {\frac{{\partial ^2 \gamma }}{{\partial I^2 }}} \right|_{I = I_R }  = \frac{{\omega ^2 \left( {\left(kc/\omega\right)^2  - 1} \right)}}{{\gamma _R }} \nonumber
\end{eqnarray}
and use the generating function $W_2=(I-I_R)\zeta+Ps^*$ to introduce new pairs of conjugate variables $(\zeta, P_\zeta)$ and $(s^*,P^*)$ with $s^*=s+(\partial I_R/\partial P)\zeta$, $P^*=P-(\partial I_R/\partial s)\zeta$:
\begin{eqnarray}
\tilde H_I  &=& \Lambda \left( {s^* ,P^* } \right) + \frac{1}{2}gP_\zeta ^2  + U_w \left( {s,I_R } \right)\sin \zeta \nonumber \\
  &\approx& \Lambda \left( {s,P} \right) + \frac{1}{2}gP_\zeta ^2  - r\zeta  + U_w \left( {s,I_R } \right)\sin \zeta \nonumber  \\
 r &=& \left\{ {\Lambda ,I_R } \right\}_{s,P}  = \frac{{\partial \Lambda }}{{\partial s}}\frac{{\partial I_R }}{{\partial P}} - \frac{{\partial \Lambda }}{{\partial P}}\frac{{\partial I_R }}{{\partial s}} \label{eq05}
\end{eqnarray}
Using Hamiltonian (\ref{eq02}), we get $m_ec^2\Delta\gamma=\omega\Delta I$ (note $\partial H_I/\partial t=0$) and $\dot I=-\partial H_I/\partial\varphi = -U_w\cos(\varphi)$. Therefore, the energy change $\Delta\gamma$ due to resonant interaction can be written as\cite{Neishtadt99, Artemyev18:cnsns}
\begin{eqnarray}
m_ec^2\Delta \gamma  &=&  - \omega U_w \int\limits_{ - \infty }^{+\infty } {\cos \varphi dt}  = - 2\omega U_w \int\limits_{ - \infty }^{t_R } {\cos \varphi dt}  \nonumber \\   = - 2\omega & U_w&  \int\limits^{ + \infty }_{\zeta _R } {\frac{{\cos \zeta }}{{gP_\zeta  }}d\zeta }
   =   - \sqrt {\frac{2}{g}} \omega \int\limits^{+\infty }_{\zeta _R } {\frac{{U_w \cos \zeta d\zeta }}{{\sqrt {h_\zeta   + r\zeta  - U_w \sin \zeta } }}}  \nonumber \\
  &= & - \sqrt {\frac{{2U_w }}{g}} \omega \int\limits^{+\infty }_{\zeta _R } {\frac{{a\cos \zeta d\zeta }}{{\sqrt {\left( {\zeta  - \zeta _R } \right) - a\left( {\sin \zeta  - \sin \zeta _R } \right)} }}}  \nonumber\\
  &=&  - \sqrt {\frac{{2r}}{g}} \omega f\left( {\zeta _R ,a} \right) \label{eq06}
\end{eqnarray}
where $t_R$ is the time of the resonant interaction, $\zeta_R$ is the value of $\zeta$ at
 $t=t_R$, and we use the Hamiltonian equation $\dot\zeta=\partial \tilde{H}_I/\partial P_\zeta$ to express $P_\zeta$ through the energy at the resonance $h_\zeta=\tilde{H}_I-\Lambda=U_w\sin\zeta_R-r\zeta_R$ (note Eq. (\ref{eq06}) is written for $r>0$, see details in Refs. \onlinecite{Neishtadt99, Artemyev18:cnsns}). Coefficient $a=U_w/r$ determines the mode of resonant interaction: for $|a|>1$ we deal with nonlinear interaction with $\langle \Delta\gamma\rangle_{h_\zeta}\ne 0$. Function $f(\zeta_R,a)$ is shown in Fig. \ref{fig2} (a). This is a periodic function with the average value \cite{Neishtadt99, Artemyev18:cnsns} equal to
\begin{equation}
\left\langle {\Delta \gamma } \right\rangle _{h_\zeta  }  =  \frac{\omega}{\pi} \sqrt {\frac{{2r}}{g}}  \int\limits_{\zeta _- }^{\zeta _+ } {\sqrt {\left( {\zeta  - \zeta _- } \right) - a\left( {\sin \zeta  - \sin \zeta _- } \right)} d\zeta }
\label{eq07}
\end{equation}
where $\zeta_\pm$ are shown in the phase portrait of $\tilde{H}_I-\Lambda$ Hamiltonian (see Fig. \ref{fig1}(b)).

The energy change in Eq. (\ref{eq07}) represents the amplitude of energy scattering $\Delta\gamma_{scat}=\left\langle {\Delta \gamma } \right\rangle _{h_\zeta  }$ and depends on the resonance position $s_R$ given by equations
\begin{equation}
h =  - \omega I_R \left( {s,P} \right) + m_e c^2 \gamma _R \left( {s,P} \right),\quad \gamma _0  = \gamma _R \left( {s,P} \right)
\label{eq08}
\end{equation}
where $m_ec^2\gamma_0$ being the initial electron energy and $h=-\omega I_{x0}+m_ec^2\gamma_0$ (with $I_{x0}$ is the initial $I_x$ value). Equations (\ref{eq08}) can be rewritten as
\begin{equation}
\gamma _R = \left| {\frac{{\Omega _{ce} }}{{kc}} \mp \frac{{kc}}{{\sqrt {\left( {kc} \right)^2  - \omega ^2 } }}\sqrt {1 + \left( {\frac{{\Omega _{ce} }}{{kc}}} \right)^2  - \frac{{2h}}{{m_e c^2 }}\frac{{\Omega _{ce} }}{{kc}}} } \right|
\label{eq09}
\end{equation}
Figure \ref{fig2}(b) shows $\Delta\gamma_{scat}(s_R)$. For a given $h$ (or equivalently for a given initial pitch-angle $\alpha_0$ determining $I_{x0}$) we can plot $\Delta\gamma_{scat}$ as a function of initial energy $m_ec^2\gamma_0$, see Figure \ref{fig2}(c). As $h$ is determined by $\alpha_{0}$ and $\gamma_0$, the amplitude of energy scattering $m_ec^2\Delta\gamma_{scat}$ depends on $\alpha_0, \gamma_0$. Analogous dependencies of scattering amplitude on initial particle parameters has been tested for several specific Hamiltonians \cite{Artemyev14:pop, Vainchtein18:jgr, Artemyev17:pre}. For Hamiltonian (\ref{eq01}) we compare the numerically calculated $\Delta\gamma_{scat}$ with the analytical expression  (\ref{eq07}) in Fig. \ref{fig2}(c): to evaluate  $\Delta\gamma_{scat}$ numerically, for several $\gamma_0$ we run $10^4$ trajectories for Hamiltonian  (\ref{eq01}) for fixed $h$ and different $\gamma_0$ (the time of integration of each trajectory includes only one resonant interaction) and then average energy changes.

In contrast to scattering, particle trapping is a non-local process. The energy change due to trapping significantly exceeds $\Delta\gamma_{scat}$ and can be comparable with the initial particle energy.  Particles can be trapped if the probability of trapping $\Pi$ is positive \cite{Neishtadt14:rms, Artemyev18:cnsns}. For system (\ref{eq05}), this probability is defined by the relation \cite{Artemyev15:pop:probability}:
\begin{equation}
\Pi  =  - \frac{{m_e c^2 }}{\omega }\frac{{d\Delta \gamma _{scat} }}{{dI}} =  - \frac{{m_e c^2 }}{\omega }\left\{ {\Delta \gamma _{scat} ,I_R } \right\}
\label{eq10}
\end{equation}
where \{$\cdot, \cdot$\}   is the Poisson bracket with respect to the variables $s,P $.  The value $\Pi$ depends on the initial energy $\gamma_0$ and $I_x$ in terms of  their combination $h=-\omega I_{x0}+m_e c^2\gamma_0$. If Eq. (\ref{eq10}) gives a negative value, $\Pi$ should be set to be zero and there are no trapped particles.

Equation (\ref{eq10}) determines the relative number of resonant particles that get trapped during a single resonant interaction. Analogous equations have been verified using the test particle calculations for several specific Hamiltonians in Refs. \onlinecite{Artemyev15:pop:probability, Artemyev17:pre, Vainchtein18:jgr}. Note that due to conservation of $h$, the change of $I$ is equal to the change of $ \gamma m_ec^2/\omega $, and thus Eq. (\ref{eq10}) can be written as $\Pi=-d\Delta\gamma_{scat}/d\gamma$, i.e. the derivative of the profile $\Delta\gamma_{scat}(\gamma)$ from Fig. \ref{fig2}(c) should give the $\Pi(\gamma)$ profile (for fixed $h$). Figure \ref{fig2}(d) shows this $\Pi(\gamma)$ and the corresponding numerical verifications (each numerical point shows the relative number of $10^4$ particles that have been trapped during the first resonant interaction).

Being trapped at some resonant value $s_R$ defined by Eqs. (\ref{eq08}), particles should escape at $s_{detrap}$ with an energy gain $\Delta\gamma_{trap}=\gamma_R(s_{detrap})-\gamma_0$. This detrapping coordinate can be calculated using the conservation of the adiabatic invariant $(2\pi)^{-1}\oint{P_\zeta d\zeta}$ for trapped particles (see details in, e.g., Refs. \onlinecite{Artemyev17:pre, Vainchtein18:jgr}). Formally speaking, $s_{detrap}$ is the solution of equation $\Delta\gamma_{scat}(s_{detrap})=\Delta\gamma_{scat}(s_R)$, i.e., the function (\ref{eq07}) should have the same value at the trapping and detrapping positions \cite{Vasiliev11, Artemyev15:pop:probability}.

To summarize, for a given $h=-\omega I_{x0}+\gamma_0 m_ec^2$, the resonant system (\ref{eq01}) can be reduced to a 1D system that is described by the profile of energy change due to scattering $\Delta\gamma_{scat}(\gamma)$, probability of trapping $\Pi=\max(0, -d\Delta\gamma_{scat}/d\gamma)$, and energy change due to trapping $\Delta\gamma_{trap}(\gamma)$. These three functions allow us to construct a map describing the system evolution on a time interval including many resonances.

\section{Mapping technique}
Let us discuss the meaning of the probability of trapping, $\Pi$. Each trajectory far from the resonance is characterized by initial energy $\gamma$, magnetic moment $I_x$, coordinates in the $(s, p_\parallel)$ plane, and phase $\zeta$ which coincides with $\varphi$. Knowing these values we can determine if particles will be trapped or scattered during the first resonant interaction. However, particle phase $\zeta$ changes with time much faster than particle $s, p_\parallel$ coordinates ($\dot\zeta \sim \Omega_{ce}$ is the largest frequency in the system). Therefore, even a small initial variation of $\zeta$ can result in a crucial change of the particle's fate: trapped particles may become scattered and vice versa. Accordingly, instead of tracing individual trajectories with given $\zeta$, it is more suitable to adopt a probabilistic approach and to consider the relative amount of trapped particle trajectories, equal to $\Pi$ (see Refs. \onlinecite{Neishtadt75, Shklyar81}).

Will a particle be trapped or scattered depends on $\zeta$ value at the resonance, $\zeta_R$, but instead of $\zeta$ it is more convenient to use the normalized resonant energy $\xi=(a\sin\zeta_R-\zeta_R)/2\pi-(a\sin\zeta_+-\zeta_+)/2\pi$ (where $\zeta_R\in[\zeta_+ - 2\pi, \zeta_- ]$, see Fig. \ref{fig1}(b)), which is distributed uniformly (see numerical tests of $\xi$ distributions in, e.g., Refs. \onlinecite{Itin00, Artemyev11:chaos}); its values belong to the interval $[0, 1]$. Within this interval, the measure of the sub-range corresponding to trapping equals $\Pi$, and we can assume for simplicity that this sub-range is $0 \leq \xi \leq \Pi$. As the particle energy does not change between two successive resonant interactions, we can write a map of the $\gamma\to\bar\gamma$ transition during a single resonance:
\begin{equation}
\begin{array}{l}
 \bar \gamma  = \gamma  + \left\{ {\begin{array}{*{20}c}
   {\Delta \gamma _{trap} \left( \gamma  \right),} & {\xi  \in \left[ {\left. {0,\Pi } \right)} \right.}  \\
   {\Delta \gamma _{scat} \left( \gamma  \right),} & {\xi  \in \left( {\left. {\Pi ,1} \right]} \right.}  \\
\end{array}} \right. \\
 \Pi  =  - d\Delta \gamma _{scat} /d\gamma  \\
 \end{array}
\label{eq11}
\end{equation}

\begin{figure}
\includegraphics[width=0.45\textwidth]{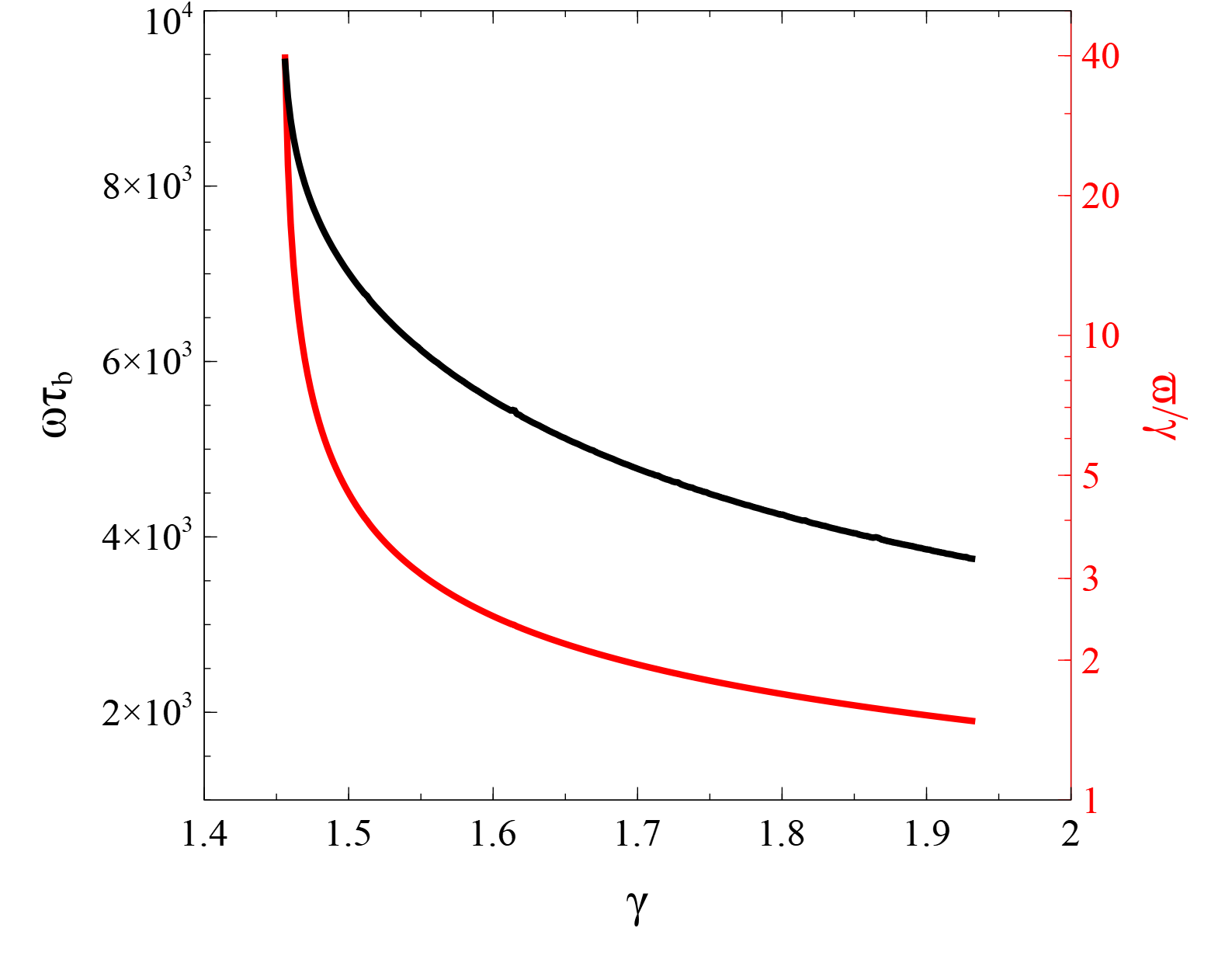}
\caption{Functions $\omega\tau_b(\gamma)$, $\varpi(\gamma)$ for fixed value of $h$ ($h/m_ec^2=1.454$ corresponding, e.g., to an equatorial pitch-angle  $\alpha_0=45^\circ$ for a $300$ keV electron; see details of model parameters in the caption of Fig. \ref{fig1}). \label{fig3}}
\end{figure}

The map (\ref{eq11}) should be supplemented with a map for phase $\xi$, which is related to $\zeta$ change (gain) between two resonances through the equation (see Appendix A):
\begin{equation}
\bar \xi=\xi-\Delta\zeta/2\pi
\label{eq11:phase}\\
\end{equation}
The rate of $\zeta$ change is defined by the Hamiltonian system (\ref{eq02}) (note that $\zeta=\varphi$), but it is more convenient to use notations of the Hamiltonian system (\ref{eq01}):
\begin{eqnarray}
 \dot \zeta  =  - \omega  + \frac{{\Omega _{ce} \left( s \right)}}{\gamma } + k\left( s \right)\frac{{p_\parallel  }}{{m_e \gamma }} \nonumber \\
\label{eq12}\\
 p_\parallel   = m_e c\sqrt {\gamma ^2  - 1 - \frac{{2I_x \Omega _{ce} \left( s \right)}}{{m_e c^2 }}}  \nonumber
\end{eqnarray}
Integrating Eq. (\ref{eq12}) over the time interval between two resonances (in the system under consideration this time is equal to the bounce period $\tau_b$), we obtain
\begin{eqnarray}
\Delta \zeta  &=& \omega \tau _b \left( {\frac{\varpi }{\gamma } - 1} \right) \label{eq13}\\
 \tau _b  &=& \frac{4}{c}\int\limits_0^{s_{\max } } {\left( {\gamma ^2  - 1 - \frac{{2I_x \Omega _{ce} \left( s \right)}}{{m_e c^2 }}} \right)^{ - 1/2} ds}  \nonumber \\
 \varpi & =& \frac{{\int\limits_0^{s_{\max } } {\Omega _{ce} \left( s \right)\left( {\gamma ^2  - 1 - \frac{{2I_x \Omega _{ce} \left( s \right)}}{{m_e c^2 }}} \right)^{ - 1/2} ds} }}{{\int\limits_0^{s_{\max } } {\omega \left( {\gamma ^2  - 1 - \frac{{2I_x \Omega _{ce} \left( s \right)}}{{m_e c^2 }}} \right)^{ - 1/2} ds} }} \nonumber
\end{eqnarray}
where $\tau_b$ and $\varpi$ depend on energy $\gamma$ and $I_x$ or, at $h$ fixed,  these functions  depend only on $\gamma$ (see Fig. \ref{fig3}). Note that the integral $\gamma^{-1}\oint{k(s)p_{\parallel} dt}=\oint{k(s)ds}$ is equal to zero.

Combining Eq. (\ref{eq13}) and Eq. (\ref{eq11}), we obtain the map for this resonant system in the $(\gamma, \zeta)$ plane:
\begin{equation}
\begin{array}{l}
 \bar \xi  = \xi  - \omega \tau _b \left( {\varpi \gamma ^{ - 1}  - 1}\right)/2\pi ,\quad \Pi  =  - d\Delta \gamma _{scat} /d\gamma \label{eq14} \\
 \bar \gamma  = \gamma  + \left\{ {\begin{array}{*{20}c}
   {\Delta \gamma _{trap} \left( \gamma  \right),} & {\bar \xi  \in \left[ {\left. {0,\Pi } \right)} \right.}  \\
   {\Delta \gamma _{scat} \left( \gamma  \right),} & {\bar \xi  \in \left( {\left. {\Pi ,1} \right]} \right.}  \\
\end{array}} \right. \\
 \end{array}
\end{equation}

\begin{figure*}
\includegraphics[width=0.95\textwidth]{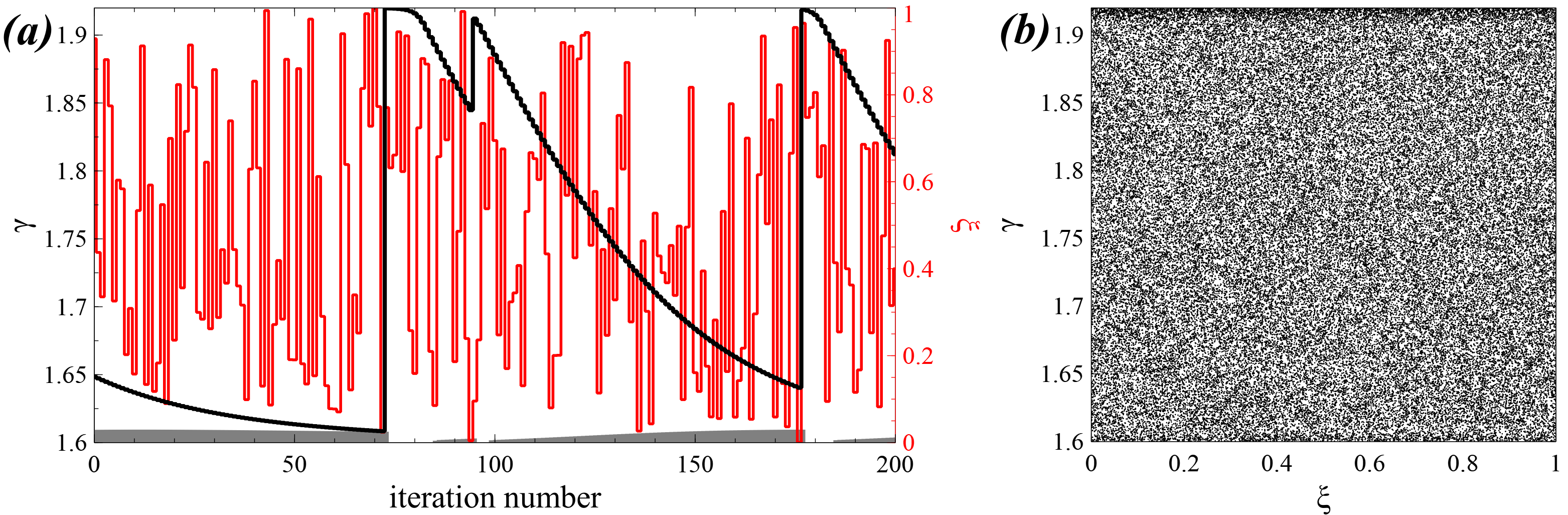}
\caption{(a) Particle trajectory for map (\ref{eq14}): $\gamma$ and $\xi$ as functions of iteration number, grey color shows the capture probability $\Pi$. (b) Particle trajectory in $(h_R, \gamma)$ plane for $10^5$ iterations. ($h/m_ec^2=1.454$ corresponds to, e.g., $\alpha_0=45^\circ$ of the equatorial pitch-angle for $300$ keV electron energy; see details of model parameters in the caption of Fig. \ref{fig1}). \label{fig4}}
\end{figure*}

This map describes variation of particle energy and phase. Figure \ref{fig4}(a) shows a typical trajectory obtained from $200$ iterations for map (\ref{eq14}): the particle loses energy due to scattering and sometimes (when the phase appears to be within the short range $[0,\Pi)$) gains energy due to trapping. After a sufficiently large number of iterations, the particle trajectory fills the entire $(\zeta,\gamma)$ plane (within the range of the resonant energies for which $\Delta\gamma_{scat}\ne0$), as shown in Fig.
\ref{fig4}(b). Such spreading of a single trajectory means that any initial distribution of energy should tend toward a uniform distribution (note that we are speaking of energies for a fixed $h$, i.e. the energy distribution along the resonant curve\cite{bookLyons&Williams, Summers98}). A similar result has been obtained through numerical simulations and solutions \cite{Mourenas18:jgr} of the kinetic equation \cite{Artemyev16:pop:letter} describing the long-term dynamics of a system consisting of many trajectories (\ref{eq01}). In the next section, to check the derived map (\ref{eq14}), we compare the results provided by this map with results of test particle simulations, as well as with results obtained by solving the full kinetic equation.

\section{Verification of mapping results}
Let us fix $h$ and consider a 1D energy distribution $\Psi(\gamma)$ (note that $\Psi$ is a cut of the 2D energy/pitch-angle distribution). We can represent this distribution as a set of $10^6$ individual particles with different initial energies and randomly distributed phases. Then, the trajectory of each particle can be traced numerically using Hamiltonian equations (\ref{eq01}) over a time interval including many resonance interactions. This method reproduces the evolution of $\Psi(\gamma)$ driven by the wave-particle resonant interaction. Alternatively, we can trace trajectories and reproduce the evolution of $\Psi(\gamma)$ using the map (\ref{eq14}). The third approach is to solve the kinetic equation that describes $\Psi$ evolution due to nonlinear resonant interactions \cite{Artemyev16:pop:letter, Artemyev18:jpp}:
\begin{equation}
\frac{{\partial \Psi }}{{\partial t}} = V\frac{{\partial \Psi }}{{\partial J }} + \frac{{dV}}{{dJ }}\left( {\Psi ^*  - \Psi } \right)\Theta \left( J \right)
\label{eq15}
\end{equation}
where $J(\gamma)=\int^{\gamma}{\tau_b(\gamma')d\gamma'}$, $V=\Delta\gamma_{scat}/\tau_b$, $\Psi^*=\Psi(\gamma^*)$ with $\gamma^*+\Delta\gamma_{trap}(\gamma^*)=\gamma$, and $\Theta=0$ for $\Pi=0$ and $\Theta=1$ for $\Pi>0$. Equation (\ref{eq15}) can be re-written in a simplified form
\begin{equation}
\frac{{\partial \Psi }}{{\partial t}} = V\frac{{\partial \Psi }}{{\partial \gamma }} + \frac{{dV}}{{d\gamma }}\left( {\Psi ^* \tau_b(\gamma^*)/\tau_b(\gamma) - \Psi } \right)\Theta \left( \gamma \right) + \ell
\label{eq16}
\end{equation}
where the term $\ell$ includes derivatives $\sim\partial\tau_b/\partial\gamma$ and can be omitted for a sufficiently weak $\tau_b(\gamma)$ dependence (see details of $\gamma\to J(\gamma)$ transformation in Refs. \onlinecite{Artemyev18:jpp, Artemyev17:arXiv}).

\begin{figure*}
\includegraphics[width=0.95\textwidth]{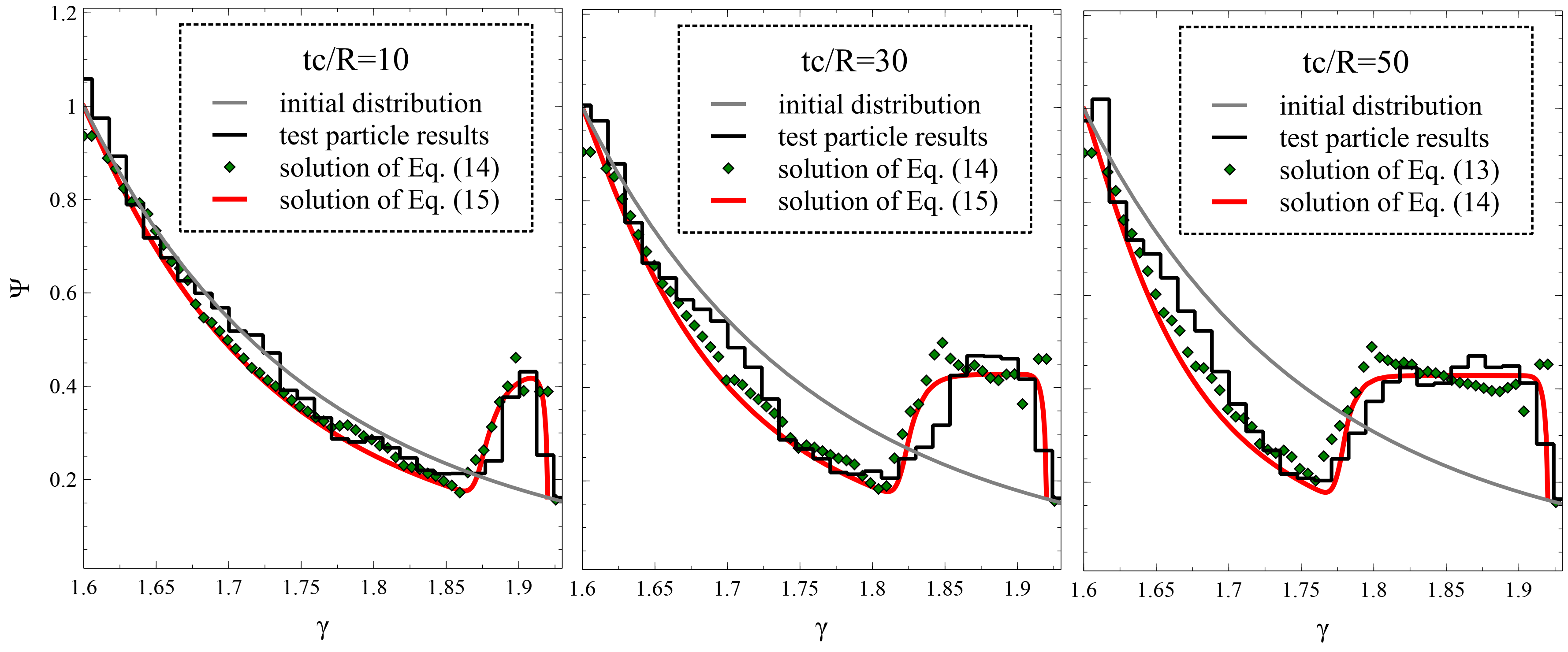}
\caption{Evolution of distribution $\Psi(\gamma)$ for $h/m_ec^2=1.454$ (this value of $h$ corresponds to, e.g., equatorial pitch-angle $\alpha_0=45^\circ$ for $300$ keV electron energy; see details of the model parameters in the caption to Fig. \ref{fig1}): black color shows results of test particle simulations ($10^6$ trajectories), red color shows solutions of Eq. (\ref{eq15}), green color shows results of mapping (\ref{eq14}). The initial distribution $\Psi(\gamma)$ is shown in all panels with grey curves. Time is normalized on $R/c$ (a scale of the quarter of the bounce period) with  $R=4.5$ Earth radii. \label{fig5}}
\end{figure*}

\begin{figure*}
\includegraphics[width=0.95\textwidth]{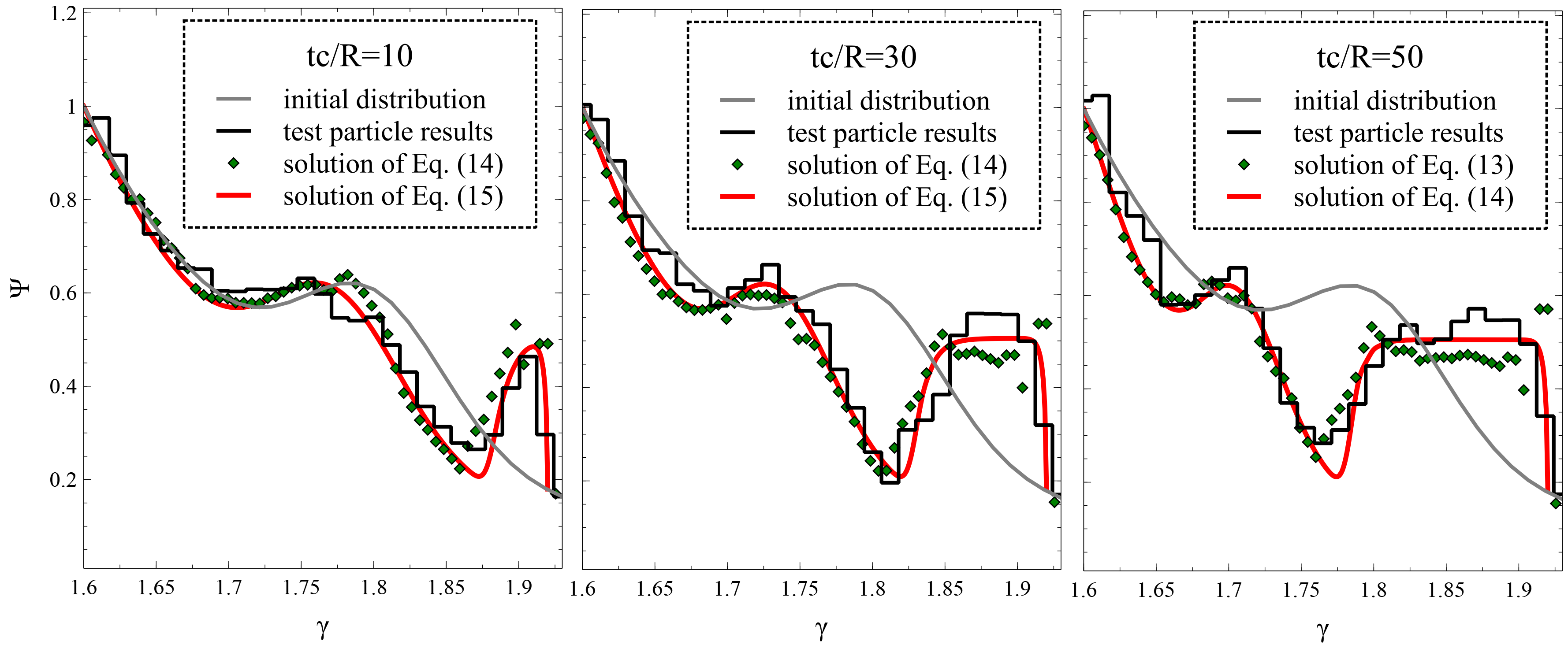}
\caption{Evolution of the distribution $\Psi(\gamma)$ for $h/m_ec^2=1.454$ (this value of $h$ corresponds to, e.g., equatorial pitch-angle $\alpha_0=45^\circ$ for $300$ keV electron energy; see details of the model parameters in the caption of Fig. \ref{fig1}): black color shows results of test particle simulations ($10^6$ trajectories), red color shows solutions of Eq. (\ref{eq15}), green color shows results of mapping (\ref{eq14}). The initial distribution $\Psi(\gamma)$ is shown in all panels with  grey curves. Time is normalized to $R/c$ (the quarter of the bounce period) with $R=4.5$ Earth radii. \label{fig6}}
\end{figure*}

We consider such three types of solutions of $\Psi(\gamma)$ evolution (test particles, Eq. (\ref{eq15}), and map (\ref{eq14})) for two initial distributions $\Psi$. Figure \ref{fig5} shows these three solutions for initial power law energy distribution and three moments of time (note that solutions obtained via test particle simulations and Eq. (\ref{eq15}) depend on time, whereas the map (\ref{eq14}) depends on the iteration number that should be transformed to time using the bounce period $\tau_b(\gamma)$ for each trajectory). All three solutions show a very similar evolution of $\Psi(\gamma)$: the distribution gets flattened within the resonant energy range (where $\Delta\gamma_{scat}\ne 0$) and forms a plateau. This is the effect of a competition between trapping (energy increase) and scattering (energy decrease), that ultimately results in a uniform distribution (note that this uniform distribution is formed along the resonance curves \cite{bookLyons&Williams, Summers98}, i.e. for $h={\rm const}$). A similar evolution, although more complicated, can be seen in Fig. \ref{fig6} showing three solutions for an initial $\Psi(\gamma)$ with a local maximum. This maximum starts drifting to lower energies (due to scattering), whereas a new maximum forms at high energies (due to trapping acceleration). Finally $\Psi(\gamma)$ will be flattened and form a plateau within the resonant energy range. Such an evolution of $\Psi(\gamma)$ has been predicted and described (considering solutions of Eq. (\ref{eq15})) in Refs. \onlinecite{Mourenas18:jgr, Artemyev19:pd}.

\section{Discussion}
In this study, we have developed a map describing the dynamics of systems with nonlinear resonant wave-particle interactions. For illustration we used wave and background plasma parameters typical for the Earth inner magnetosphere, where relativistic electrons resonate with high amplitude whistlers. This system is well investigated in the regime of low wave amplitudes where quasi-linear diffusion equations are applicable \cite{bookLyons&Williams, Shprits08:JASTP_local, Thorne10:GRL}, but so far there is no method allowing to model the long-term evolution of this system in the presence of nonlinear resonant effects. One of the most widespread technique, test particle simulation \cite{Demekhov06, Omura07, Katoh&Omura07:acceleration, Bortnik08, Demekhov09, Mourenas16, Agapitov16:grl}, provides a lot of important information about electron acceleration and scattering rates, but such simulations are limited to quite short time intervals. This limitation mostly comes from the need to integrate the entire (bounce) particle trajectory even if energy and magnetic moment only change at the locations of wave-particle resonances. Therefore, a natural solution consist in considering only resonance-induced changes of particle energy and pitch-angle, like  in the quasi-linear diffusion approach. The generalization of the diffusion equation with inclusion of terms related to nonlinear wave-particle interaction results in Eq. (\ref{eq15}) or similar types of kinetic equations \cite{Solovev&Shkliar86, Artemyev16:pop:letter, Artemyev18:jpp}. However, this kinetic equation still  relies on the assumption of a uniform distribution of resonant phases (i.e.,  it excludes effects related to phase correlation at multiple passages through the resonance) and it cannot be easily generalized for systems with multiple waves. These two problems can be resolved using the map approach that includes resonant phase dynamics while also allowing the inclusion of many resonances. Let us consider these two issues in more details.

Assuming a uniform distribution of resonant phases corresponds to the assumption that two successive resonances are not correlated, i.e., that electron energy jumps $\Delta \gamma$ (due to trapping or scattering) can be considered as independent over a long run. This important property of the resonant system usually results from the dependence of the phase gain $\Delta\zeta$ on energy $\gamma$ (see Eq. (\ref{eq14})). This gain is usually large $\Delta\zeta\sim \omega\tau_b\gg 1$ (since whistler wave period is much smaller than the electron bounce period along magnetic field lines) and, thus, even a small change of energy $\Delta \gamma$ due to resonant interaction should result in a significant change of $\Delta\zeta$: $\delta\left(\Delta\zeta\right)\sim \left(\partial\Delta\zeta/\partial\gamma\right)\Delta \gamma$, justifying the assumption of randomly distributed phases. However, resonances can be correlated (and the distribution of resonant phases can be non-uniform, see Ref. \onlinecite{Artemyev19:arXiv}) for systems with small $\partial\Delta\zeta/\partial\gamma$. Such a situation can hardly appear in the Earth radiation belts, but it is more realistic for resonant electron interaction with strong electrostatic waves and solitons around the bow shock \cite{Vasko18:grl}. This corresponds to Landau resonant interaction without the term $\varpi/\gamma$ in Eq. (\ref{eq14}) and with the time interval between resonances $\sim \tau_b$ weakly depending on energy. Therefore, the proposed map technique may be useful for investigations of nonlinear wave-particle interactions in such systems, where the assumption of a uniform distribution of resonant phases is not applicable.

The map (\ref{eq14}) has been constructed for a system with a single wave (single resonance). In this system the condition $h={\rm const}$ reduces the initially 2D space (energy/pitch-angle or $(\gamma, I_x)$) to 1D space. However, unlike kinetic equation (\ref{eq15}), this map can be generalized to many resonances resulting in particle motion in the $(\gamma, I_x)$ space. Indeed, the map describes change of the resonant phase $\zeta$ between two resonances and energy change on the resonance. The $\zeta$ change can be modified by replacing the integration over the entire bounce period with the integration between two resonances in Eq. (\ref{eq13}), whereas the energy change can be replaced with energy and $I_x$ changes. This generalization looks much simpler to achieve than the corresponding generalization of the kinetic equation (\ref{eq15}).

\begin{figure}
\includegraphics[width=0.45\textwidth]{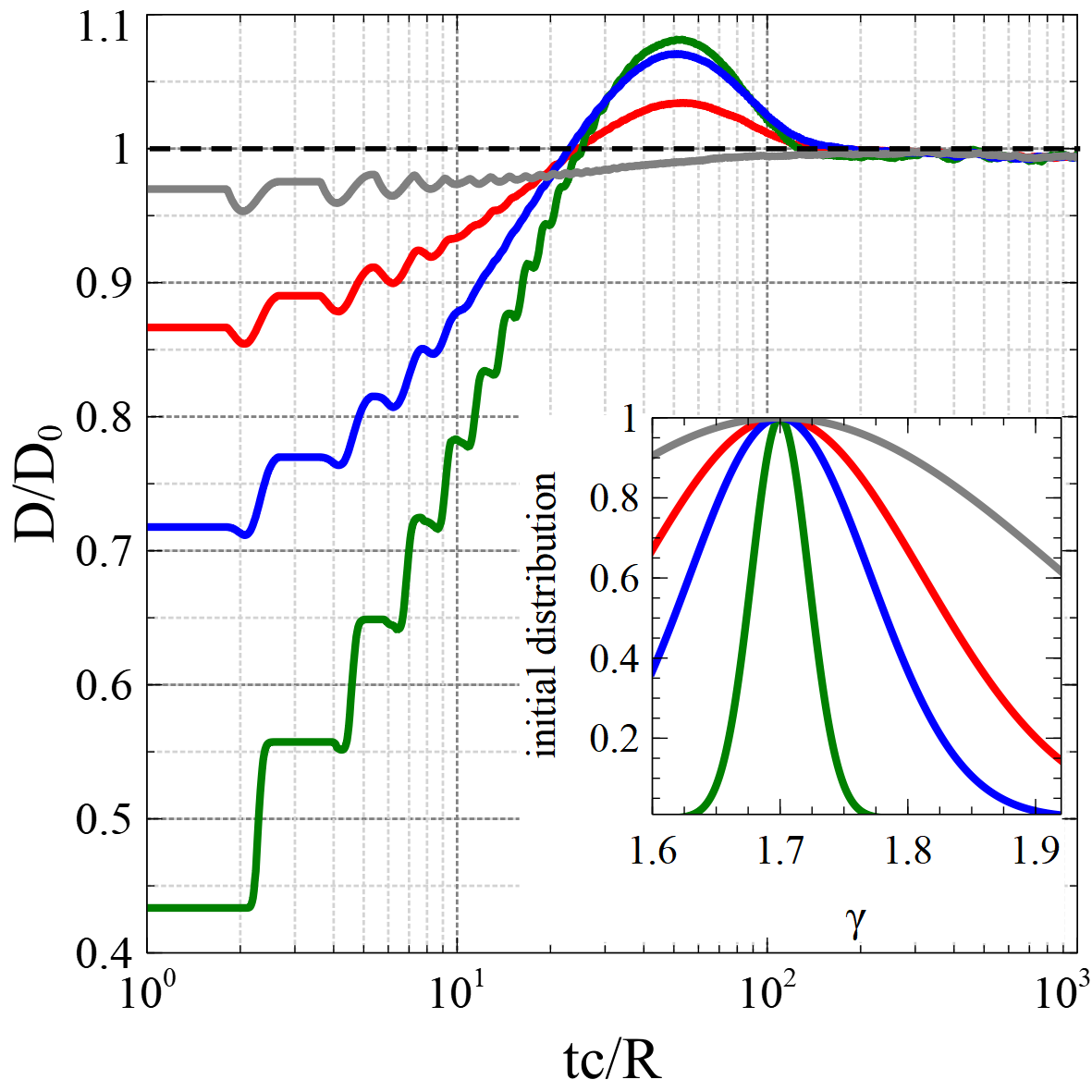}
\caption{Evolution of dispersion $D=\sqrt{\langle\gamma^2\rangle-\langle\gamma\rangle^2}$ for four distributions $\Psi(\gamma)$ and $h/m_ec^2=1.454$ (this value of $h$ corresponds to, e.g. $45^\circ$ of the equatorial pitch-angle for $300$ keV electron energy; see details of model parameters in the caption of Fig. \ref{fig1}). Dispersions are normalized to the dispersion $D_0$ of the uniform distribution $\Psi={\rm const}$. \label{fig7}}
\end{figure}

Figure \ref{fig4} shows that after many iterations the particle trajectory fills the entire available space in the $(\zeta,\gamma)$ space. For ensembles including many trajectories, the final state of the distribution function will be a uniform distribution where phase space density $\Psi$ should have the same value for all energies. This is the final state for both quasi-linear diffusion, that tends to reduce gradients of $\Psi$ along the resonance curve, and nonlinear wave-particle interaction described by Eq. (\ref{eq15}), which has only one stationary solution $\Psi={\rm const}$ (see Ref. \onlinecite{Artemyev19:pd}). Therefore, the map (\ref{eq14}) describes distribution flattening, $\partial \Psi/\partial\gamma\to 0$, and allows to estimate a typical  timescale of this flattening. For example, Fig. \ref{fig7} shows the evolution of the dispersion $D=\sqrt{\langle\gamma^2\rangle-\langle\gamma\rangle^2}$ of distribution $\Psi$ for four different initial $\Psi$ (shown in the inserted panel). The dynamics of $\Psi$ is described by $10^5$ trajectories of map (\ref{eq14}), and $D$ is normalized to the dispersion of the uniform distribution for the same $\gamma$ range (i.e., $D/D_0\approx 1$ means $\Psi\approx {\rm const}$). As the map (\ref{eq14}) describes discrete changes of energy with time, the beginning of the $D$ evolution consists of step-like jumps (note that we transformed the iteration number to time for each trajectory to plot $D$ versus time). Independently of the initial distribution, $D/D_0\approx 1$ after $\approx 100\tau_b$, and this timescale is much shorter than typical quasi-linear time scales \cite{Thorne13:nature, Li13}.

To conclude, in this paper we have considered nonlinear resonances between relativistic electrons and intense whistler-mode waves. We have demonstrated that the long-term dynamics of the electron distribution can be described by a map taking into account the important interdependence between the probability of trapping $\Pi$ and energy change due to scattering $\Delta \gamma_{scat}$: $\Pi=-d \Delta\gamma_{scat}/d\gamma$. This map is different from the classical Chirikov map \cite{Chirikov79}, and allows to describe both effects of phase trapping and nonlinear scattering. The proposed mapping technique can be useful for the description of charged particle acceleration in various space plasma systems including the Earth radiation belts and the Earth bow shock.

\begin{acknowledgments}
Authors are grateful to Dr. D. Mourenas for useful discussion. This work was supported by the Russian Scientific Fund, project 19-12-00313.
\end{acknowledgments}

\section*{Appendix A}
In this Appendix we derive the relation between $h_R$ change between two resonance crossings and phase $\zeta$ gain between these crossings. We consider quite general form of Hamiltonian, but results of our derivations are directly applicable to the particular system considered in the main text. 

Let us consider a general Hamiltonian as a sum of unperturbed part $H_0(I, p,q)$ and small perturbation $\varepsilon H_1(I,\zeta, p,q)$ (with $\varepsilon \ll 1$) where  $(q, \varepsilon^{-1} p)$, $(\zeta, I)$ are pairs of conjugate variables: hence $(\zeta, I)$ are fast variables, $(q, p)$ are slow variables, and $H_1$ is periodic in $\zeta$ (note in the main text $\varepsilon\sim B_w/B_0$). Momentum $I$ is the adiabatic invariant: $I$ is constant in the unperturbed system $H=H_0$, whereas under the perturbation $\dot I=-\varepsilon\partial H_1/\partial \zeta$.
There is no explicit dependence on time, and thus $H=h={\rm const}$. The resonance condition is determined by the equation $\partial H_0/\partial I=0$ that gives $I=I_R(p,q)$. We introduce $\Lambda(p,q)=H_0(I_R(p,q),p,q)$ and the improved adiabatic invariant $J$ with the variable transformation $(I,\zeta, p,q)\to (J,\theta, P,Q)$, such that new Hamiltonian is $H=H_0(J, P, Q)+\varepsilon\bar H_1(J, P,Q)$ (in the leading approximation) with $\bar H_1=\langle H_1 \rangle_\zeta$ (note new $H=h={\rm const}$).

\begin{figure}
\includegraphics[width=0.45\textwidth]{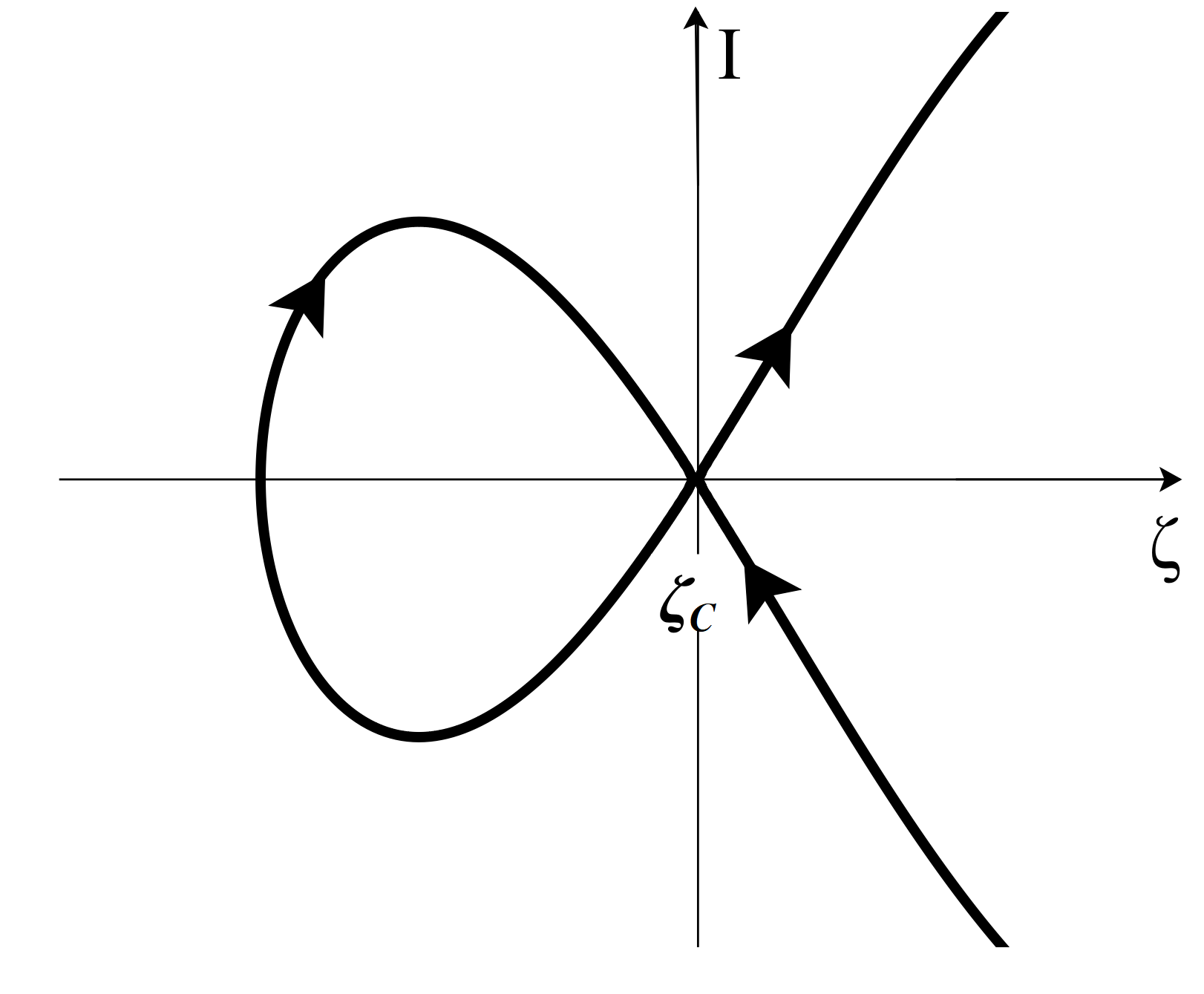}
\caption{Schematic of phase portrait. \label{figA}}
\end{figure}

We consider the resonance surface $I=I_R$ and assume that near this surface the phase portrait of the original Hamiltonian looks like one shown in Fig. \ref{figA}.
Far from the resonance $\theta$ changes
with the frequency
\begin{equation}
\dot \theta  = \frac{{\partial H_0 }}{{\partial J}} + \varepsilon \frac{{\partial \bar H_1 }}{{\partial J}}
\label{eq01A}
\end{equation}
with $J={\rm const}$, and
\begin{equation}
\dot Q = \varepsilon \frac{{\partial H_0 }}{{\partial P}} + \varepsilon ^2 \frac{{\partial \bar H_1 }}{{\partial P}},\quad \dot P =  - \varepsilon \frac{{\partial H_0 }}{{\partial Q}} - \varepsilon ^2 \frac{{\partial \bar H_1 }}{{\partial Q}}
\label{eq02A}
\end{equation}

We introduce $\omega_0(J,P,Q)=\partial H_0/\partial J$, $\omega_1(J,P,Q)=\partial \bar H_1/\partial J$, and consider large number $N\gg 1$ of $\zeta$ rotations from $t=t_0$ (when phase point is far from the resonance and moves towards the resonace) to $t=t_N$;  the last turn is sufficiently far from the resonance and $\theta \approx \zeta$ in the leading approximation. The last turn ends at $\zeta=\zeta_c \ {\rm mod}\ 2\pi$ (see Fig. \ref{figA} for $\zeta_c$ definition). Then
\begin{equation}
\zeta _{cN}  + 2\pi N = \zeta _0  + \int\limits_{t_0 }^{t_N } {\left( {\omega _0  + \varepsilon \omega _1 } \right)dt}
\label{eq04A}
\end{equation}
where $\zeta_{cN}$ is $\zeta_c$ value at $t=t_N$. We introduce $t_*$ as  the time of crossing  the resonance, i.e. $\omega_0(J, P, Q)|_{t=t_*}=0$ and rewrite Eq. (\ref{eq04A}):
\begin{equation}
\zeta _{cN}  + 2\pi N = \zeta _0  + \int\limits_{t_0 }^{t_* } {\left( {\omega _0  + \varepsilon \omega _1 } \right)dt}  - \int\limits_{t_N }^{t_* } {\left( {\omega _0  + \varepsilon \omega _1 } \right)dt}
\label{eq05A}
\end{equation}
Because time interval $t_*-t_N\ll 1/\varepsilon$ we can use $\dot Q=\varepsilon \partial H_0/\partial P$, $\dot P=-\varepsilon \partial H_0/\partial P$ and $Q\approx q$, $P\approx p$ in the last integral in Eq. (\ref{eq05A}). We also assume that $\zeta_{cN}\approx \zeta_{c*}=\zeta_c|_{t=t_*}$. To describe system dynamics for $t\in [t_N,t_*]$ we use the expansion of the Hamiltonian around the resonance $H=\Lambda+F$ and
\begin{equation}
F = \frac{1}{2}g\left( {I - I_R } \right)^2  + \varepsilon H_1 ,\;\;
 g = \left. {\frac{{\partial ^2 H_0 }}{{\partial I^2 }}} \right|_{I = I_R }  \approx {\rm const} \label{eq06A}
\end{equation}
The Hamiltonian in new variables $(J,\theta, P, Q)$ can be expanded as
\begin{equation}
H = \Lambda  + \frac{1}{2}g\left( {J - I_R } \right)^2  + \varepsilon \bar H_1
\label{eq07A}
\end{equation}
We introduce $\bar e=g(J-I_R)^2/2$ and write
\begin{eqnarray}
\dot {\bar e} &=& \varepsilon g\left( {J - I_R } \right)r,\quad r =  - \left\{ {I_R ,\Lambda } \right\} \approx {\rm const} \nonumber\\
 \omega _0  &=& \frac{{\partial H}}{{\partial J}} = g\left( {J - I_R } \right) \label{eq08A}
\end{eqnarray}
Using $dt= d\bar e/(d\bar e/dt)$ we rewrite integral
\begin{equation}
\int\limits_{t_N }^{t_* } {\left( {\omega _0  + \varepsilon \omega _1 } \right)dt}  \approx \frac{1}{\varepsilon }\int\limits_{\bar e_N }^0 {\frac{{\omega _0 d\bar e}}{{g\left( {J - I_R } \right)r}}}  = \frac{1}{\varepsilon }\int\limits_{\bar e_N }^0 {\frac{{d\bar e}}{r}}  =  - \frac{{\bar e_N }}{{\varepsilon r}}
\label{eq09A}
\end{equation}
where we take into account that $\bar e_*=0$ (by definition) and omit $\varepsilon \omega_1$ because $t_*-t_N\ll 1/\varepsilon$. Using definition $\bar e+\varepsilon \bar H_1=F$, we write
\begin{equation}
\bar e_N  = F_N  - \varepsilon \bar H_1  \approx F_N  - \varepsilon \bar H_{1*}
\label{eq10A}
\end{equation}
where $\bar H_{1*}$ is the resonant value of $\bar H_1$. Substituting Eqs. (\ref{eq09A}, \ref{eq10A}) to Eq. (\ref{eq05A}), we obtain
\begin{equation}
\zeta _{c*}  + 2\pi N = \zeta _0  + \int\limits_{t_0 }^{t_* } {\left( {\omega _0  + \varepsilon \omega _1 } \right)dt}  + \frac{{F_N  - \varepsilon \bar H_{1*} }}{{r\varepsilon }}
\label{eq11A}
\end{equation}
or
\begin{equation}
\frac{{F_N }}{{2\pi r\varepsilon }} = \frac{{\bar H_{1*} }}{{2\pi r}} - \frac{1}{{2\pi }}\left( {\zeta _0  - \zeta _{c*}  + \int\limits_{t_0 }^{t_* } {\left( {\omega _0  + \varepsilon \omega _1 } \right)dt} } \right)\,\,\, \bmod 1
\label{eq12A}
\end{equation}
We define $F_{last}$ as an $F$ value at the last crossing $\zeta=\zeta_c$ before the resonance. Thus, $F_{last}=F_N \ {\rm mod}\ 2\pi \varepsilon r$, because the change of $F$  after one $\zeta$ rotation equals to $2\pi\varepsilon r$. We introduce $\xi=(F_{last}-\varepsilon H_{1c*})/(2\pi\varepsilon r)$ (where $H_{1c*}$ equals to $H_{1}$ at $\zeta=\zeta_c$, $t=t_*$) and $\tau=\varepsilon t$, and write
\begin{equation}
\xi  = {\mathop{\rm Frac}\nolimits} \left( {\frac{{\bar H_{1*}  - H_{1c*} }}{{2\pi r}} - \frac{{\zeta _0  - \zeta _{c*}}}{{2\pi }}  - \int\limits_{\tau _0 }^{\tau _* } {\frac{{\left( {\omega _0  + \varepsilon \omega _1 } \right)}}{{2\pi\varepsilon }}d\tau }  } \right)
\label{eq13A}
\end{equation}
Note that $\xi$ can be written as
\begin{eqnarray}
\xi  &=& \frac{{F_{last}  - \varepsilon H_{1c*} }}{{2\pi \varepsilon r}} = \frac{{F_{last}  - \varepsilon r\zeta _{c*}  - \left( {\varepsilon H_{1c*}  - \varepsilon r\zeta _{c*} } \right)}}{{2\pi \varepsilon r}} \nonumber\\
&=& \frac{{E_{last}  - E_{c*} }}{{2\pi \varepsilon r}}
\label{eq14A}
\end{eqnarray}
where
\begin{equation}
E = \frac{1}{2}g\left( {I - I_R } \right)^2  - \varepsilon r\zeta  + \varepsilon H_1,
\label{eq15A}
\end{equation}
and $E_{c*}$ is $E$ value at $\zeta=\zeta_c$, $t=t_*$, $E_{last}$ is $E$ value at last crossing of $\zeta=\zeta_c$ before resonance. 
Therefore, change of $\xi$ between two resonance crossings at moments $\tau_*$ and $\tau_{*+}$ equals to the difference $\Delta\xi=\xi(\tau_{*+})-\xi(\tau_{*})$ given by Eq. (\ref{eq13A}), i.e. $\xi$ gain is
\[
\Delta \xi=-\int\limits_{\tau _* }^{\tau _{*+} } {\frac{{\left( {\omega _0  + \varepsilon \omega _1 } \right)}}{{2\pi\varepsilon }}d\tau },
\]
and this is the phase $\zeta$ gain between two resonance crossings normalized on $2\pi$ (between moments $\tau_*$ and $\tau_{*+}$) taken with the minus sign.

\bibliographystyle{elsarticle-harv}

\end{document}